\documentclass[twocolumn,showpacs,preprintnumbers,amssymb,floatfix]{revtex4}

\usepackage{graphicx}
\usepackage{dcolumn}
\usepackage{bm}

\def \be  {\begin{equation}}
\def \ee  {\end{equation}}
\def \ba  {\begin{eqnarray}}
\def \ea  {\end{eqnarray}}
\def \baa {\begin{eqnarray*}}
\def \eaa {\end{eqnarray*}}
\def \bb  {\begin {thebibliography} }
\def \eb  {\end{thebibliography}}
\def \lab #1 {\label{#1}}

\newcommand \bi [1] {\bibitem{#1}}
\newcommand\re[1]{(\ref{#1})}

\def \matrix #1 {\left(\begin{array}{cc} #1 \end{array}\right)}

\def \Im {\mathop{\rm Im}\nolimits}
\def \Re {\mathop{\rm Re}\nolimits}

\def \e  {\mathop{\rm e}\nolimits}
\newcommand\lr[1]{{\left({#1}\right)}}
\newcommand \widebar [1] {\overline{#1}}

\newcommand{\as}{\ifmmode\alpha_{\rm s}\else{$\alpha_{\rm s}$}\fi}
\newcommand{\asbar}{\ifmmode\bar{\alpha}_{\rm s}\else{$\bar{\alpha}_{\rm s}$}\fi}

\font\cmss=cmss12 
\def\inbar{\,\vrule height1.5ex width.4pt depth0pt}
\def\IC{\relax\hbox{$\inbar\kern-.3em{\rm C}$}}
\def\IZ{\relax{\hbox{\cmss Z\kern-.4em Z}}}
\def\IR{{\hbox{{\rm I}\kern-.2em\hbox{\rm R}}}}

\def\IP{{\hbox{{\rm I}\kern-.2em\hbox{\rm P}}}}
\def\II{\hbox{{1}\kern-.25em\hbox{l}}}

\begin{document}

\preprint{\parbox{30mm}{LPT--Orsay--01--109 \\ TPJU--12/2001 \\ UB--ECM--PF--01/13 \\
hep-ph/0111185}}


\title{Solution of the multi-reggeon compound state problem in multi-colour QCD}

\author{G.P.~Korchemsky${}^1$}
\author{J.~Kota{\'n}ski${}^2$}
\author{ A.N.~Manashov${}^3$}
\affiliation{
${}^1$~Laboratoire de Physique Th\'eorique, Universit\'e de Paris XI, 91405
Orsay, France, \\ ${}^2$~Institute of Physics, Jagellonian University, Reymonta
4, PL-30-059 Cracow,
Poland\\
${}^3$~Department d'ECM, Universitat de Barcelona, 08028 Barcelona, Spain}

\date{November 15, 2001 }

\begin{abstract}
We study the properties of the colour-singlet compound states of reggeized gluons
in multi-colour QCD using their relation with noncompact two-dimensional
Heisenberg spin magnets. Applying the methods of integrable models, we calculate
their spectrum and discuss the application of the obtained results to high-energy
asymptotics of the scattering amplitudes in perturbative QCD.
\end{abstract}

\pacs{12.38.-t,04.20.Jb,03.65.Db,12.40.Nn}

\maketitle

{\bf 1.}~High-energy asymptotics of the scattering amplitudes in perturbative QCD
are dominated by the contribution of an infinite number of soft gluons exchanged
in the $t-$channel between the scattered particles. Due to the remarkable
property of gluon reggeization, it becomes possible to resum the corresponding
Feynman diagrams by formulating an effective theory, in which collective gluonic
excitations -- the reggeized gluons, or reggeons, plays the r\^ole of new
elementary fields \cite{L2}. In this way, the scattering amplitudes describe the
propagation of the reggeons in the $t-$channel and their interaction with each
other. Each diagram in the effective theory becomes equivalent to an infinite sum
of the Feynman diagrams involving ``bare'' gluons. To the leading logarithmic
approximation, $\as\ln s={\rm fixed}$ as $s\to\infty$, the dominant contribution
comes from the diagrams with only two reggeons exchanged in the $t-$channel. It
leads to rising of the scattering amplitudes $A(s,t)$ with the energy $s$ -- the
BFKL Pomeron \cite{L2}, but violates the unitarity constraints at high energies.
To restore the unitarity of $A(s,t)$ in the direct ($s-$, $t-$ and $u-$)
channels, one has to take into account the $t-$channel exchange of an arbitrary
number ($N=3, 4, ...$) of reggeons. For fixed $N$, their contribution to $A(s,t)$
satisfies the BKP equation \cite{BKP}. Its solution defines the colour-singlet
compound states built from $N$ reggeons
and satisfying the Schr\"odinger equation\\[-3mm]
\begin{equation}
{\cal H}_N \Psi(\vec z_1,\vec z_2,...,\vec z_N) = E_N \Psi(\vec z_1,\vec
z_2,...,\vec z_N)
\label{Sch}
\end{equation}
\\[-3mm]
with the effective QCD Hamiltonian ${\cal H}_N$ acting on two-dimensional
transverse coordinates of reggeons, $\vec z_k$  ($k=1,...,N$) and their colour
$SU(N_c)$ charges. In the large $N_c$ limit the colour structure of the
Hamiltonian simplifies significantly leading to ${\cal H}_N=\sum_{k=1}^N H(\vec
z_k,\vec z_{k+1})$ with $\vec z_{N+1}\equiv \vec z_1$ and $H(\vec z_k,\vec
z_{k+1})$ being the BFKL kernel \cite{L2}.

High-energy asymptotics of the scattering amplitude due to the $t-$channel
exchange of the compound $N-$reggeon states is given by
\\[-4.2mm]
\be
A(s,t) \sim -is\sum _{N=2}^\infty (i \asbar)^N \frac{s^{-\asbar E_N/4}}{(\asbar
\ln s)^{1/d_N}}\,\xi_{1,N}(t)
\xi_{2,N}(t)
\label{amp}
\ee
\\[-4mm]
with $\asbar=\as N_c/\pi$, $E_N$ being the energy of the ground state for
\re{Sch} and the exponent $d_N$ parameterizing the distribution of the energy
levels next to the ground state (see Eq.~\re{accum} below). The residue factors
$\xi_{1(2),N}(t)$ measure the overlap of $\Psi(\vec z_1,\vec z_2,...,\vec z_N)$
with the wave functions of two scattered particles. At $N=2$ and $N=3$ the
corresponding compound states are known as the BFKL Pomeron \cite{L2} and the
Odderon \cite{LN,JW}, respectively.

In this letter, we shall present
the results of our calculations of the ground state energy $E_N$ for higher
reggeized gluon compound states in multi-colour QCD and discuss their physical
interpretation.

{\bf 2.}~Our approach to solving the Schr\"odinger equation \re{Sch} is based on
the remarkable properties of integrability of the effective QCD Hamiltonian
${\cal H}_N$ in the multi-colour limit \cite{L1,FK}. It turns out that, for
$N_c\to\infty$, the Schr\"odinger equation \re{Sch} possesses the hidden set of
the conserved charges $[{\cal H}_N,q_k]=[q_k,q_m]=0$ and is completely
integrable. It defines a quantum-mechanical model of $N$ interacting particles on
the two-dimensional plane, which can be identified as two-dimensional noncompact
Heiseinberg spin chain model \cite{DKM}. The number of sites of the model is
equal to the number of reggeons, $N$, and the spins attached to each site,
$S_k^{\pm,0}$ and $\bar S_k^{\pm,0}$, are six generators of the principal series
of the $SL(2,\mathbb{C})$ group. Using the (anti)holomorphic coordinates on the
two-dimensional $(x,y)-$plane, $z=x+iy$ and $\bar z=x-iy$, one represents the
spin operators as the differential operators $S^0=z\partial_z +j$,
$S^-=-\partial_z$, $S^+=z^2\partial_z+2jz$. The operators $\bar S^{\pm,0}$ are
given by similar expressions with $z$, $j$ replaced by $\bar z$, $\bar j$,
respectively. The pair of complex parameters $(j,\bar j)$ specifies the
$SL(2,\mathbb{C})$ representation. For the principal series of the
$SL(2,\mathbb{C})$ they are parameterized by integer $n_j$ and real $\nu_j$ as
$j=(1+n_j)/2+i\nu_j$ and $\bar j=1-j^*$. To match the reggeon Hamiltonian,
Eq.~\re{Sch}, the spins have to be equal to $j=0$ and $\bar j=1$. In this
representation, the BFKL kernel defining the interaction between two reggeons
describes the interaction between the nearest spins
\\[-5mm]
\be
{\cal H}_N = \sum_{k=1}^N \left[ H(J_{k,k+1}) + H(\bar J_{k,k+1}) \right],
\label{Ham}
\ee
\\[-3mm]
where $H(J)=\psi(1-J)+\psi(J)-2\psi(1)$ with $\psi(x)=d\ln\Gamma(x)/dx$,
$J_{k,k+1}$ is the sum of two $SL(2,\mathbb{C})$ spins,
$J_{k,k+1}(J_{k,k+1}-1)=(\vec S_k+\vec S_{k+1})^2$ with $J_{N,N+1}\equiv
J_{N,1}$, and similar for $\bar J_{k,k+1}$.
Eq.~\re{Ham} implies that the dynamics in the $z-$ and $\bar z-$sectors is
independent on each other. The interaction between two sectors is ensured by the
condition that the eigenstates $\Psi(\vec z_1,\vec z_2,...,\vec z_N)$ have to be
single-valued functions on the two-dimensional plane.

{\it The two-dimensional noncompact spin magnet\/} defined by Eqs.~\re{Sch} and
\re{Ham} can be solved exactly within the $R-$matrix approach by using the method
of the Baxter $\mathbb{Q}-$operator. As was shown in \cite{DKM}, the Hamiltonian
is expressed in terms of a single operator $\mathbb{Q}(u,\bar u)$
\\[-4.5mm]
\begin{eqnarray}
{\cal H}_N&=&i\frac{d}{du} \ln \left\{ u^{2N}
\left[\mathbb{Q}(u-i(1-j),u-i(1-\bar j))\right]^\dagger\right.
\nonumber \\[-1mm]
&&\left.\left.\phantom{\frac{}{}}\times\mathbb{Q}(u+i(1-j),u+i(1-\bar
j))\right\}\right|_{u=0}
\label{H-Q}
\end{eqnarray}
\\[-4mm]
and the Schr\"odinger equation \re{Sch} turns out to be equivalent to the
eigenproblem for the Baxter operator. Its eigenvalues, $Q(u,\bar u)$, satisfy the
following conditions \cite{DKM}.
Firstly, $Q(u,\bar u)$ has to fulfil 
the chiral Baxter equation
\begin{equation}
t_N(u)\,Q(u,\bar u)\!=\! (u+ij)^N\,Q(u+i,\bar u) + (u-ij)^N\,Q(u-i,\bar u)
\label{Bax-eq}
\end{equation}
with $t_N(u)=2u^N+q_2 u^{N-2} + ... +q_N$ and  $q\equiv(q_2,\ldots,q_N)$ being
the eigenvalues of the holomorphic integrals of motion. $Q(u,\bar u)$ obeys
similar equation in the $\bar u-$sector with $\bar j=1-j^*$ and $\bar q_k=q_k^*$.
The lowest integral of motion, $q_2$, is related to the total $SL(2,\mathbb{C})$
spin of the system, $h$,
\be
q_2=-h(h-1)+Nj(j-1)\,,\quad h={(1+n_h)}/2+i\nu_h
\label{q2}
\ee
with $n_h$ integer and $\nu_h$ real. Secondly, at $u=\lambda-in/2$ and $\bar
u=\lambda+in/2$ with $n$ integer, $Q(u,\bar u)$ has to be an analytic function on
the complex $\lambda-$ plane except the points $\{u_{m}^{\pm}=\pm i(j-m)\,, \bar
u_{\bar m}^{\pm}=\pm i(\bar j-\bar m)\}$, with $m,\bar m$ positive integer, where
it has $N$th order poles. The behaviour of $Q(u,\bar u)$ at the vicinity of the
pole at $m=\bar m=1$ can be parameterized as \cite{MW,DKM,DL}\\[-5mm]
\be
Q(u_{1}^{\pm}+\epsilon,{\bar u}_{1}^{\pm}+\epsilon)=R^\pm(q)\left[
\frac1{\epsilon^N} +\frac{i\,E^\pm(q)}{\epsilon^{N-1}}+ ... \,\right]\!.
\label{Q-R,E}
\ee
\\[-3mm]
The functions $R^\pm(q)$ fix normalization of the Baxter operator, while the
residue functions $E^\pm(q)$ define the energy of the system (see Eq.~\re{energy}
below).

Important properties of the solutions to \re{Sch} can be deduced from the
symmetry of the Hamiltonian \re{Ham} under the cyclic permutations of particles,
$\mathbb{P}\Psi(\vec z_1,...,\vec z_{N-1},\vec z_N)\!=\!\Psi(\vec z_2,...,\vec
z_N,\vec z_1)$ and their mirror permutations $\mathbb{M}\Psi(\vec z_1,...,\vec
z_{N-1},\vec z_N)\!=\!\Psi(\vec z_N,...,\vec z_2,\vec z_1)$ \cite{DKM}. The
eigenstate of the Hamiltonian diagonalizes the operator $\mathbb{P}$ and the
corresponding eigenvalue defines its quasimomentum
\be
\mathbb{P}\,\Psi_q(\vec z_i)=\e^{i\theta_{N}(q)}\Psi_q(\vec z_i)\,,\qquad \theta_N(q)=2\pi{k}/{N}
\label{quasi}
\ee
with $k$ integer in virtue of $\mathbb{P}^N=1$. The operator $\mathbb{M}$ acts on
the integrals of motion as $q_k\to (-1)^k q_k$ and maps the eigenstate of the
Hamiltonian into another one with the same energy but different set of the
quantum numbers, $\mathbb{M}\Psi_q=\Psi_{-q}$ with $-q\equiv(q_2,-q_3,...,(-1)^n
q_n)$, leading to
\begin{equation}
\!E_N(q)=E_N(-q),~~
\label{E-parity}
Q_{q}(-u,-\bar u) = \e^{i\theta_{N}(q)} Q_{-q}(u,\bar u).
\end{equation}
Inserting \re{Q-R,E} into \re{E-parity}, we find that $E^\pm(q)=-E^\mp(-q)$.
Then, substituting \re{Q-R,E} into \re{H-Q} and using hermiticity of the
Hamiltonian \re{Ham}, we obtain the energy as \cite{DKM}
\be
E_N(q)
=\Re\left[E^+(-q)+ E^+(q)\right].
\label{energy}
\ee

{\bf 3.}~Determining the eigenvalues of the Baxter operator, $Q(u,\bar u)$, we
use the following ansatz~\cite{FK}
\\[-4mm]
\be
Q(u,\bar u)= \int{d^2 z}\, z^{-i u-1} {\bar z}^{-i\bar u-1}\, Q(z,\bar z),
\label{Q-R}
\ee
\\[-3mm]
where $Q(z,\bar z)$ is a single-valued function on the 2-dimen\-sional $\vec
z-$plane with $\bar z=z^*$. The Baxter equation \re{Bax-eq} leads to the $N$-th
order differential equation on $Q(z,\bar z)$
\\[-3mm]
\ba
&&
\left[z^j\lr{z\partial_z}^{N}z^{1-j}+z^{-j}\lr{z\partial_z}^{N}z^{j-1}
-2\lr{z\partial_z}^{N}
\right]Q(z,\bar z)
\nonumber
\\[-2mm]
&&=\sum_{k=2}^N i^{k}q_k\lr{z\partial_z}^{N-k} Q(z,\bar z)
\label{Eq-1}
\ea
\\[-3mm]
The $\bar z-$dependence of $Q(z,\bar z)$ is fixed by a similar equation in the
$\bar z-$sector with $\bar j=1-j^*$, $\bar q_k=q_k^*$.

The differential equation \re{Eq-1} has three regular singular points: $z=0$,
$z=1$ and $z=\infty$. The solutions to \re{Eq-1} around $z=\infty$ can be
obtained from those around $z=0$, using the symmetry of \re{Eq-1} under the
transformation $z\to 1/z$ and $q_k\to (-1)^k q_k$. The solutions to \re{Eq-1} at
the vicinity of $z=0$ and $z=1$ have the form $Q\sim\,z^a$ and $\sim\,(1-z)^b$,
respectively. Solving the indicial equations one finds that the exponents $a$ are
$N-$times degenerate, $(a-1+s)^N=0$, while $b$ are given by $b_1=Nj-h-1$,
$b_2=Nj+h-2$ and $b_m=m-3$ for $m=3,\ldots,N$. Since the exponents differ from
each other by an integer, this may lead to the terms $\sim\ln z$ and $\sim
\ln(1-z)$. This is the case at $z=0$, whereas at $z=1$ the solutions to \re{Eq-1}
do not contain terms $\sim\ln(1-z)$ unless $h=(1+n_h)/2$.

Following \cite{JW}, we define the fundamental set of solutions to Eq.~\re{Eq-1},
$Q_m^{_{(0)}}(z)$, $(m=1,...,N)$, around $z=0$ as
\\[-4mm]
\be
Q_m^{(0)}(z)=z^{1-j}\sum_{k=0}^{m-1}c_{m-1}^k u_{k+1}(z) (\ln z)^{m-k-1}\,,
\label{Q-0-h}
\ee
\\[-3mm]
with the coefficients $c_{m-1}^k=\Gamma(m)/(\Gamma(k+1)\Gamma(m-k))$ chosen for
later convenience. The functions $u_m(z)$ are given by the power series $u_m(z) =
1+\sum_{k=1}^\infty z^k\,u_{m,k}$ with the coefficients $u_{m,k}=u_{m,k}(q)$
satisfying the three-term recurrence relations with respect to $k$. Defining
similar set of the fundamental solutions in the $\bar z-$sector, we construct the
general solution for $Q(z,\bar z)$ around $z=0$ as
\\[-4mm]
\be
Q(z,\bar z) = \sum_{m,\bar m=1}^N Q^{(0)}_m(z)\, C^{(0)}_{m\bar
m}\,\widebar{Q}^{(0)}_{\bar m}(\bar z)
\label{Q-0}
\ee
\\[-3mm]
with $C^{(0)}$ being the mixing matrix. The fundamental solutions
$Q^{_{(0)}}_m(z)$ and $\widebar{Q}^{_{(0)}}_{\bar m}(\bar z)$ have a nontrivial
monodromy around $z=\bar z=0$. For $Q(z,\bar z=z^*)$ to be single-valued at
$z=0$, the monodromy should cancel in the r.h.s.\ of \re{Q-0}. To satisfy this
requirement, $C^{_{(0)}}_{nm}$ should vanish for $m\ge N-n$, and have the
following form for $m \le N+1-n$
\be
C^{(0)}_{nm}=
\sum _{k=0}^{N-n-m+1}{\frac {(-2)^{k}\,\alpha_{k+n+m-1}}{k!(n-1)!(m-1)!}}
\label{C0}
\ee
with $\alpha_1,...,\alpha_N$ being arbitrary complex parameters. To fix the
normalization of the function $Q(z,\bar z)$ we choose $\alpha_N=1$. Substituting
\re{C0} and \re{Q-0-h} into \re{Q-0} we find
\\[-4mm]
\be
\!Q(z,\bar z)=z^{1-j}\bar z^{1-\bar j}\!\left[\frac{\ln^{N-1} (z\bar z)}{(N-1)!} +
\frac{\ln^{N-2} (z\bar z)}{(N-2)!}\, \alpha_{N-1}+\!...\right]\!\!,
\label{Q-small}
\ee
\\[-5mm]
where ellipses denote the terms subleading at $z=0$.

As follows from \re{Q-R}, the small$-z$ asymptotics of the function $Q(z,\bar z)$
is in one-to-one correspondence with the singularities of $Q(u,\bar u)$.
Substituting \re{Q-small} into \re{Q-R} and performing integration over $|z|\ll
1$ one finds that, in agreement with \re{Q-R,E}, $Q(u,\bar u)$ has the $N$th
order pole at $u_1^+=i(j-1)$ and $\bar u_1^+=i(\bar j-1)$
\\[-3mm]
\be
Q(u_{1}^{+}+\epsilon,{\bar u}_{1}^{+}+\epsilon)=-\frac{\pi}{(i\epsilon)^N}
\left\{1 + i\epsilon\alpha_{N-1}
+{\cal O}(\epsilon^2)
\right\}\!.
\label{Q-pole-0}
\ee
\\[-3mm]
Matching this expression into \re{Q-R,E}, one obtains $E^+(q)=\alpha_{N-1}(q)$.
This leads to the following remarkably simple expression for the energy
\re{energy}
\be
E_N(q)=\Re\left[\alpha_{N-1}(-q)+\alpha_{N-1}(q)\right],
\label{E-fin}
\ee
where $q=\{q_k\}$ and $-q=\{(-1)^k q_k\}$. Thus, the energy is related to the
small$-z$ asymptotics of the function $Q(z,\bar z)$, Eq.~\re{Q-small}, or
equivalently to the matrix elements of the mixing matrix
\re{C0} in the fundamental basis \re{Q-0-h}.

The fundamental set of solutions to Eq.~\re{Eq-1} around $z=1$ is defined
similarly to \re{Q-0-h} as
\be
Q_m^{(1)}(z) = z^{1-j} (1-z)^{b_m} v_m(z)
\label{Q-1-h}
\ee
with $m=1,...,N$. Here, the functions $v_n(z)$ are given by the power series,
$v_n(z)\!=\!1+\sum_{k=1}^\infty (1-z)^k v_{n,k}$ with the expansion coefficients
$v_{n,k}$ satisfying the $N-$term recurrence relations with respect to $k$. To
fix uniquely their solutions, we impose the normalization conditions
$v_m(z)=1+{\cal O}((1-z)^{N-m+1})$ for $m\ge 3$. As was already mentioned, at
$h\!\!=\!\!(1+n_h)/2$ the  solutions $Q_{1,2}^{_{(1)}}(z)$ become degenerate and
one of them, $Q_1^{_{(1)}}(z)$ for $n_h\ge 0$, has to be redefined by including
the additional $\ln(1-z)-$term
\\[-5mm]
\ba
Q_1^{(1)}(z)\!\!\!\!&&
= z^{1-j} (1-z)^{Nj-(n_h+3)/2}
\nonumber\\
&&\times\left[(1-z)^{n_h}
\ln (1-z)\, v_2(z)+ v_1(z)\right]\,.
\label{Q-deg}
\ea
\\[-5mm]
Using \re{Q-1-h} and similar expressions in the $\bar z-$sector, one constructs
the general solution to \re{Eq-1} for $\Im h\neq 0$ that possesses a trivial
monodromy around $z=1$ as
\\[-5mm]
\ba
&&\hspace*{-12mm}Q(z,\bar z)=\beta_1 Q_1^{(1)}(z)\widebar Q_1^{(1)}(\bar
z)+\beta_2 Q_2^{(1)}(z)\widebar Q_2^{(1)}(\bar z)
\nonumber
\\[-1mm]
&&\hspace*{-7mm}+\sum_{m,\bar m=3}^N Q_m^{(1)}(z)\,\gamma_{m\bar m}\,\widebar
Q_{\bar m}^{(1)}(\bar z)\equiv
\overrightarrow{\!Q}{}^{(1)}\!
\cdot C^{(1)}\!
\cdot\overrightarrow{\!\!\widebar Q}{}^{(1)}.
\label{Q-1}
\ea
\\[-4mm]
At $h=(1+n_h)/2$, or equivalently $\Im h=0$, the first term in the r.h.s.\ of
\re{Q-1} looks differently in virtue of \re{Q-deg}
\be
\!Q(z,\bar z)=\beta_1\!\left[Q_1^{(1)}(z)\widebar Q_2^{(1)}(\bar
z)+Q_2^{(1)}(z)\widebar Q_1^{(1)}(\bar z)\right]+\! ...,
\ee
where ellipses denote the remaining terms.

Eqs.~\re{Q-0} and \re{Q-1} provide the solutions to the Baxter equation at the
vicinity of $z=0$ and $z=1$, respectively. Choosing $z$ to be inside the region
of convergence of the both expressions, we define the transition matrices
\\[-5mm]
\be
Q_n^{(0)}(z)=\sum_{m=1}^N \Omega_{nm}\, Q_m^{(1)}(z)
\label{Omega-def}
\ee
\\[-4mm]
and similar for $\widebar\Omega$. For the fundamental set of solutions,
Eqs.~\re{Q-0-h} and \re{Q-1-h}, these matrices are uniquely fixed. They can be
calculated as $\Omega=W^{(0)}[W^{(1)}]^{-1}$, with
$W^{_{(j)}}_{kn}=\partial^n Q_k^{_{(j)}}(z_0)$ and $z_0$ being some reference
point~\cite{JW}.

The transition matrices allow us to analytically continue the solutions \re{Q-0}
from $z=0$ to $z=1$. Substituting \re{Omega-def} into \re{Q-0} and matching the
resulting expression for the function $Q(z,\bar z)$ into \re{Q-1}, we find the
relation between the mixing matrices $C^{(0)}$ and $C^{(1)}$
\be
C^{(1)}=\left[\Omega(q)\right]^T C^{(0)}\,\widebar
\Omega(\bar{q})\,.
\label{C1-C0}
\ee
This matrix equation provides the quantization conditions for the integrals of
motion $q$ and fixes uniquely the eigenvalues of the Baxter operator. Replacing
the mixing matrices by their expressions, Eqs.~\re{Q-0} and \re{Q-1}, we obtain
from \re{C1-C0} the system of $N^2$ equations involving $(N-1)$ parameters
$\alpha_n$ inside $C^{(0)}$, $2+(N-2)^2$ parameters $\beta_{1,2}$ and
$\gamma_{m\bar m}$ inside $C^{(1)}$, as well as $(N-2)$ integrals of motion
$q_3,...,q_N$ (we recall that $\bar q_k=q_k^*$). Thus, the system \re{C1-C0} is
overcomplete. Not only it allows to determine all parameters including the
quantized $q$, but it also provides $(2N-3)$ nontrivial consistency conditions.

{\bf 4.}~Following this approach, it becomes straightforward to solve  the
Schr\"odinger equation \re{Sch} for arbitrary $SL(2,\mathbb{C})$ spins $(j,\bar
j)$ and any finite $N$. In what follows, we shall present the results of our
calculations of the {\it ground state energy\/} of \re{Sch} at $j=0$ and $\bar
j=1$. Its value defines the energy of the $N-$reggeon states, $E_N$, entering
Eq.~\re{amp}. The detailed description of the full spectrum will be presented in
the forthcoming publication.

At $N=2$, the equation \re{Eq-1} can be solved exactly and the general expression
for the function $Q(z,\bar z)$ be written in terms of the Legendre
$\mathbf{Q}-$functions of the 2nd kind
\\[-4mm]
\be
\!\!\!Q(z,\bar z)\! =\!\frac{z}{(1-z)^{2}}\!\left[Q_h(z) \widebar Q_{\bar h}(\bar z)\! -
\!Q_{1-h}(z)\widebar Q_{1-\bar h}(\bar z)\right]\!,
\label{Q-N=2}
\ee
\\[-3mm]
with $Q_h(z)\equiv\mathbf{Q}_{-h}\left((1+z)/(1-z)\right)$. Its expansion 
at $z=0$ and $z=1$ matches Eqs.~\re{Q-small} and \re{Q-1}, respectively, and
leads to $\alpha_1(q)=\alpha_1(-q)=E_2(h)/2$ with
\be
E_2(h)=4\Re\left[\psi(h)+\psi(1-h)-2\psi(1)\right]
\label{N=2}
\ee
and $\e^{i\theta_2}=(-1)^{n_h}$, in agreement with the known exact expression
\cite{L2}. The ground state is unique and it corresponds to $h=1/2$, $E_2=-16\ln
2$ and $\theta_2=0$.
\begin{table}[t]
\begin{center}
\begin{tabular}{|c||c|c|c|c|c||r|r|}
\hline
  &   $q_2$ & $ iq_3$ &  $q_4$ & $iq_5$ &  $q_6$ & $-E_N/4$
  & $\sigma_N/4\,\,\,$
\\
\hline
${N=2}$ & .25 &           &           &           &
& 2.77259
&16.8288
\\
${N=3}$
& .25
& .20526 
&           &           &
& -.24717 
&.9082
\\
${N=4}$
& .25
&  0
& .15359 
&           &
& .67416 
& 1.3176
\\
${N=5}$
& .25
& .26768 
& .03945 
& .06024 
&
& -.12751 
& .4928
\\
${N=6}$
& .25 &         0
& .28182 
& 0
& .07049 
& .39458 
& .5634
\\
\hline
\end{tabular}
\end{center}
\vspace*{-4mm}
\caption{Quantum numbers, $q$, and the energy, $E_N$, of the
$N-$reggeon states in multi-colour QCD. The parameter $\sigma_N$ defines the
energy of the lowest excited states,  Eq.~\re{accum}.}
\label{tab:energy}
\vspace*{-5mm}
\end{table}

For $N\!\ge\!3$, the analytical solution to \re{Eq-1} is not available and one
has to rely on the power series solutions, Eqs.~\re{Q-0-h} and \re{Q-1-h}.
Performing thorough analysis of the quantization conditions \re{C1-C0} at
$N=3,4,5,6$ with a help of the method described in \cite{KP} (see Appendix C), we
obtained the results summarized in the Table~\ref{tab:energy}. At $N=3$ they are
in agreement with \cite{JW}.

The quantized values of the integrals of motion have a rich structure -- they
form the family of one-dimensional continuous trajectories in the
$(N\!\!-\!1)$-dimensional space of $q\!\!=\!\!(q_2,...,q_N)$. The ``proper time''
along each trajectory is defined by the imaginary part, $\nu_h$, of the conformal
spin $h=(1+n_h)/2+i\nu_h$. The total number of trajectories is infinite. Each of
them is labelled by the set of integers including the quasimomentum
$\theta_N=2\pi k/N$ and the parameter $n_h$ that defines the total 2-dimensional
angular momentum of the $N-$reggeon state. In an agreement with our physical
intuition, the ground state of the system belongs to the trajectory with
$\theta_N=0$ and $n_h=0$. The energy along this trajectory is a continuous
function of $\nu_h$ and it approaches its minimal value (=ground state energy
$E_N$) at $\nu_h=0$. At the vicinity of $\nu_h=0$, one finds an accumulation of
the energy levels,
\\[-5mm]
\be
E_N(\nu_h)-E_N =\sigma_{N} |\nu_h|^{d_N},
\label{accum}
\ee
\\[-5mm]
with $d_N=2$ and the coefficients $\sigma_N$ given in the Table~\ref{tab:energy}.

The ground state of the $N-$reggeon system has different properties for even and
odd $N$. For even $N$, the ground state is {\it unique\/} and it belongs to the
trajectory, on which the quantized $q_3,...,q_N$ satisfy the relations $\Im
q_{2k+2}=q_{2k+1}=0$ with $k=1,...,(N-2)/2$. For odd $N$, the ground state is
{\it double degenerate\/}. The corresponding eigenstates belong to two different
trajectories, on which the quantum numbers satisfy the relations $\Im
q_{2k+2}=\Re q_{2k+1}=0$ and $\Im q_{2k+1}\neq 0$ with $k=1,...,(N\!-\!1)/2$. In
distinction with even $N$, quantized $q_3,q_5,...$ do not vanish and take pure
imaginary values on the ground state trajectories.
%
%
Two ground state trajectories differ from each other only by the sign of
$q_{2k+1}$ and the degeneracy of the ground state occurs due to the symmetry of
the energy under $q_{2k+1}\!\to\!-q_{2k+1}$, Eq.~\re{E-parity}.
The 
reason for this is \cite{DKM} that at odd $N$ there exist two mutually orthogonal
ground states, $\Psi_q^{_{(\pm)}}(\vec z_i)=[\Psi_q(\vec z_i)\pm\Psi_{-q}(\vec
z_i)]/2$, which are invariant under the cyclic permutations, $\theta_N=0$, and
possess a definite parity under the mirror permutations,
$\mathbb{M}\,\Psi_q^{_{(\pm)}}(\vec z_i)\!=\!\pm\Psi_q^{_{(\pm)}}(\vec z_i)$. In
contrast, at even $N$ one has $\Psi_{-q}(\vec z_i)=\Psi_{q}(\vec z_i)$ due to
$q_{2k+1}\!=\!0$ and, as a consequence, $\Psi_q^{_{(+)}}=\Psi_{q}$ and
$\Psi_q^{_{(-)}}=0$.

In high-energy QCD, in virtue of the Bose symmetry, the compound states of $N$
reggeized gluons has to be symmetric under interchange of any pair of reggeon
coordinates and their colour indices. In the multi-colour limit, this symmetry is
reduced to the invariance of the wave function under the cyclic and mirror
permutations. The fact that the wave function of the state is factorized, as
$N_c\!\!\to\!\!\infty$, into the product of the colour tensor and the scalar
function $\Psi_q^{(\pm)}(\vec z_1,...,\vec z_N)$ implies that the both factors
have to possess the same parity under the
cyclic and mirror permutations. This allows to distinguish the degenerate ground
states according to their $C-$parity. The $N-$gluon states possessing the parity
$\mathbb{M}=(-1)^N$ have the same $C-$parity as the BFKL Pomeron, $C=1$. For odd
$N$, the ground states with the parity $\mathbb{M}=1$ have the $C-$parity of the
Odderon state \cite{LN}, $C=-1$.

Solving the system 
\re{C1-C0}, we find the mixing coefficients
and calculate the ground state energy from \re{E-fin}. The results are presented
in the last two columns in Table~\ref{tab:energy}. For $E_N<0$ ($E_N>0$) the
contribution of the $N-$reggeon state to the scattering amplitude $A(s,t)/s$,
Eq.\re{amp}, increases (decreases) at high energy $s$. Our results indicate that
for {\it even\/} $N$ the ground state energy, $E_N$, is {\it negative\/} and it
increases with $N$ approaching the value $E_{2\infty}=0$ from below. For {\it
odd\/} $N$, the ground state energy is {\it positive\/} and it decreases with $N$
approaching the same value $E_{2\infty +1}=0$ from above. This suggests that the
ground state has the properties of an antiferromagnet.

We conclude that, in the multi-colour limit, in the Pomeron sector, only states
with even $N$ provide the contribution to \re{amp} rising with the energy $s$.
Their intercept, however, is much smaller than that of the BFKL Pomeron. In the
Odderon sector, the corrections to the ``bare''  Odderon $N=3$ state due to
higher states built from odd number of reggeons, $N$, become important. They
increase the effective value of the Odderon intercept and lead to
$\alpha_{_{\mathbb{O}}}=1$. Finally, in the both sectors, the contribution of the
$N-$reggeon states to the scattering amplitude ceases to depend on the energy $s$
as $N\to\infty$.

We are most grateful to S.{\'E}.\ Derkachov for collaboration at the early stage
and to J.\ Wosiek for useful discussions. This work was supported by the grants
KBN-PB-2-P03B-19-17 (J.K.) and RFFI-0-1-5-00 (A.M).
\\[-8mm]

\end{document}